\documentclass{aa}

\input{psfig.tex}
\def\gsim{\;\lower.6ex\hbox{$\sim$}\kern-7.75pt\raise.65ex\hbox{$>$}\;}
\def\lsim{\;\lower.6ex\hbox{$\sim$}\kern-7.75pt\raise.65ex\hbox{$<$}\;}
\begin{document}

%   \thesaurus{06     % A&A Section 6: Form. struct. and evolut. of stars
%	      (08.08.1;    %Stars: abundances --
%               08.05.3;    %Stars: evolution --
%               08.16.3;    %Stars: Population II --
%	       10.07.2)}   %Galaxy: globular cluster
%

%
\title{ Abundances of C, N, O in slightly evolved stars in the globular clusters
NGC 6397, NGC 6752 and 47 Tuc\thanks{Based on data collected at the European 
Southern Observatory, Chile (ESO LP 165.L-0263)} }

\author{
E. Carretta\inst{1},
R.G. Gratton\inst{2},
S. Lucatello\inst{2},
A. Bragaglia\inst{1}
\and
P. Bonifacio\inst{3}
}

\authorrunning{E. Carretta et al.}
\titlerunning{Abundance of C,N,O in slightly evolved globular cluster stars}

%          \fnmsep\thanks{Just to show the usage
%          of the elements in the author field}

\offprints{E. Carretta, carretta@pd.astro.it}

\institute{
INAF - Osservatorio Astronomico di Bologna, Via Ranzani 1, I-40127
 Bologna, Italy
\and
INAF - Osservatorio Astronomico di Padova, Vicolo dell'Osservatorio 5, I-35122
 Padova, Italy
%\and
%Dipartimento di Astronomia, Universit\`a di Padova, Italy
\and
INAF - Osservatorio Astronomico di Trieste, via Tiepolo 11, Trieste, Italy
%\and
%ESO - European Southern Observatory
  }

\date{Received ; accepted }

\abstract{Abundances of C and N are derived from features due to the CH G-band 
and to the UV CN band measured on high resolution ($R\gsim 40,000$) UVES
spectra of more than 40 dwarfs and subgiants in NGC 6397, NGC 6752 and 47 Tuc. 
Oxygen abundances (or upper limits) are available for all stars in the sample.
Isotopic ratios $^{12}$C/$^{13}$C were derived from the CH molecular band. This
is the first determination of this ratio in unevolved dwarf stars in globular
clusters. By enlarging the sample of subgiants in NGC 6397 studied in Gratton
et al. (2001), we uncovered, for the first time, large variations 
in both Na and O also in this cluster. The origin of the chemical
inhomogeneities must be searched for outside the stars under scrutiny.  Our data
indicate that in unevolved or slightly evolved stars in these clusters C
abundances are low but not zero, also in stars with large N-enhancements and
O-depletions, and that the isotopic ratios $^{12}$C/$^{13}$C are low, but never
reach the equilibrium value of the CN-cycle. When coupled with the run of O and
Na abundances, these findings possibly require that, in addition to CNO
burning and $p-$captures, some  triple$-\alpha$ process is also involved:
previously evolved intermediate-mass AGB stars are then the most likely
polluters.
}

%\email{
%carretta@pd.astro.it,
%gratton@pd.astro.it
%}

\maketitle

\section{INTRODUCTION}

After H and He, C, N, O are, together with Ne, the most abundant elements in
the Universe. As such, they are key ingredients in a large number of
astrophysical issues. Their abundances in metal-poor stars are tracers of the
nucleosynthetic sites that contributed to the different phases of galactic
evolution. Moreover, they are important contributors to the opacity in stellar
interiors and act as catalysts in the CNO-cycle of H-burning.

Stars of globular clusters (GCs) offer an ideal diagnostic in order to
understand
stellar evolution for low and intermediate stellar masses.
However, since the pioneering study of Osborn (1971) it is known that a spread
in the light elements (C, N, O, but also among heavier species such as Na, Al and
Mg) is present among cluster stars of similar evolutionary phase, unlike their
analogs in the galactic field (e.g. Gratton et al. 2000 and references
therein). Among these last, C and N abundances follow well defined
evolutionary paths, with two episodes of mixing, the first one related to the
first dredge-up and the second one after the red giant branch (RGB) bump, when
the molecular weight barrier created by the maximum inward penetration of the
outer convective envelope is canceled by the outward expansion of the
H-burning shell. The first episode was theoretically predicted by Iben (1964);
while the second one is not present in canonical non-rotating models, it
may be easily accomodated in models where some type of circulation is
activated e.g. by core rotation (Sweigart \& Gross 1978; Charbonnel 1994).
Gratton et al. (2000) showed that among field stars no variations 
corresponding to these mixing episodes are observed for the remaining
elements (namely, O and Na): again, this agrees with models, that do not allow
deep enough mixing along the RGB. Clearly this pointed toward a peculiarity of
globular cluster stars.
 
As shown by Denisenkov \& Denisenkova (1989), and later in a more
quantitative way by Langer et al. (1993), the observed star-to-star scatter
in cluster stars may be explained by the CNO-cycle and the accompanying
proton-capture reactions at high temperature. More uncertain is where these
reactions occurred, whether in the observed star themselves, prior to an
internal (extra- or enhanced-) very deep mixing episode, or elsewhere, perhaps
in some form of H-burning at high temperature taking place e.g. in now extinct
intermediate-mass AGB stars (IM-AGB), followed by ejection of polluting matter
(see Gratton, Sneden and Carretta 2004 for an updated review and references on
the huge literature on this subject).

Evidences from red giants are ambiguous, since both mixing (causing a decrease
of [C/Fe] as a function of luminosity, see Bellman et al. 2001 and references
therein) and pollution/accretion of processed matter (e.g. Yong et al. 2003,
Sneden et al. 2004) might be invoked to explain observations.

Cleaner conclusions can be drawn from unevolved or slightly evolved stars,
where no mixing is expected and inner temperatures are not high enough to
permit the $p-$capture reactions in the NeNa and AlMg cycles. However, up to
a short time ago, only low-dispersion observations of molecular bands of
hydrides such as CH and NH or bi-metallic molecules such as CN were available to study
the chemical composition of faint GC turn-off (TO) dwarfs or subgiants (SGB).
More important, no O indicator was accessible, since the atmospheric cutoff
and the low throughput of existing spectrographs severely hampered the use of
OH bands in the UV regions, and the remaining O features are only observable
on high dispersion spectra.

In spite of these limitations, a number of studies (see e.g. Briley et al.
2004b and references therein) uncovered that large spreads in the CH and CN
band strengths, anticorrelated with each other, do exist in unevolved stars in
several GCs. The only explanation must necessarily rest on an event that
polluted the material forming these stars, likely early in the cluster
lifetime.

Previous results from the present program (Gratton et al. 2001, Carretta et al.
2004) provided further strong evidences favouring primordial abundance
variations. Deriving the first reliable O abundances in cluster dwarfs, we
found a clear anticorrelation between Na and O among TO and SGB stars in NGC
6752 and 47 Tuc. In NGC 6397, early results were not conclusive, but they were
hampered by small number statistics. Similar anticorrelations were found also
for Mg and Al. These facts require some non internal mechanism.

However, after initial successes (see e.g. Ventura et al. 2001), more recent
models of metal-poor IM-AGB stars met serious problems in reproducing the O-Na
anticorrelation and related phenomenology (Denissenkov \& Herwig 2003; Fenner
et al. 2004; Herwig 2004), and unveiled that not only Hot Bottom Burning
occurs (HBB), but also vigorous H-burning at somewhat cooler temperatures
during the interpulse phases. To overcome these problems, Denissenkov \& Weiss
(2004) recently proposed that the site for the $p-$capture reactions is the
interior of RGB stars slightly more massive than those currently observed in
globular clusters, and that they exchanged mass with the currently unevolved
stars where the anomalous abundances are observed.

In the present paper we complete the analysis of the spectra presented in
Gratton et al. (2001) and Carretta et al. (2004)  by including detailed
abundances of C, N and O. From these abundances and
from isotopic $^{12}$C/$^{13}$C ratios, measured for the
first time in such unevolved stars, we suggest that both
triple-$\alpha$ captures in He-burning, to form fresh $^{12}$C, and typical
H-burning processing at high temperatures are required to reproduce the
observed pattern of abundances in these stars. If confirmed, this would
exclude the possibility that mass-exchange with RGB stars might be responsible
for the observed abundances. In the discussion, we will also comment on other
possible shortcomings of this hypothesis.

Furthermore, we also present results for Na and O abundances in a more
extended sample of subgiants in NGC 6397, showing that large variations,
anticorrelated with each other, in these two elements do exist also in this
metal-poor cluster.

\section{OBSERVATIONS}

Details of observations are given in Gratton et al. (2001; Paper I)  and
Carretta et al. (2004). Briefly, spectra were acquired using the
Ultraviolet-Visual Echelle Spectrograph (UVES) mounted at the ESO VLT-UT2
within several runs (June and September 2000; August and October 2001, July
2002) of the ESO Large Program 165.L-0263 (P.I. R. Gratton). On the whole, we
have observational material for 6 dwarfs and 9 subgiant stars in NGC 6397, 9
dwarfs and 9 subgiants in NGC 6752 and 3 dwarfs and 9 subgiants in 47 Tuc.
Relevant data for the observed stars are given in Gratton et al. (2001) and
Carretta et al. (2004). Those for the additional subgiants in NGC 6397 are 
the same as given for the other subgiants in Gratton et al. (2001).

Data were acquired using the dichroic beamsplitter \#2. In the blue arm we
used the CD2, centered at 420 nm, to cover both the  CH G-band at $\sim
4300$~\AA\ and the CN UV system at $\sim 3880$~\AA. The spectral coverage is
about $\lambda\lambda$ 356-484 nm. The CD4, centered at 750 nm (covering
$\lambda\lambda$ 555-946 nm), was adopted in the red arm. Observations in the
run of June 2000 (mostly for NGC 6397) were made with a slightly different
setup, resulting in a spectral coverage  $\lambda\lambda$ 338-465 nm in the
blue and $\lambda\lambda$ 517-891 nm in the red. Slit length was always 8
arcsec, while the slit width was mostly set at 1 arcsec (corresponding to a
resolution of 43000). In a few cases, according to the seeing conditions, this
value  was slightly modified downward or upward.

In NGC 6397 and NGC 6752, typical exposure times were $\sim 1$ hour for
subgiants, while in 47 Tuc we doubled this time. Each turn-off star was
observed for a total of about 4 hours, split into several exposures. 
At $\lambda \sim 4300$ we reached a typical value of the $S/N$ per
pixel of $\sim 30$, increasing to $\sim 70$ for stars in NGC 6397.

\section{ATMOSPHERIC PARAMETERS}

The adopted values of the atmospheric parameters are discussed in detail in
Gratton et al. (2001) and Carretta et al. (2004). Here we only recall the main 
features of the analysis, that was also applied to the 6 newly observed SGB
stars in NGC 6397.

We compared effective temperatures from observed colours (both Johnson $B-V$
and Str\"omgren $b-y$ were used) with spectroscopic temperatures derived from
fitting Balmer lines (namely H$\alpha$). This approach was devised to derive
precise values of the reddenings on the same scale for both cluster stars and
field stars. These were used in the estimate of accurate distances to these
clusters (see Gratton et al. 2003).

Average values of temperatures were finally used for stars of NGC 6752 and NGC
6397. However, in 47 Tuc we found that the adoption of individual T$_{\rm
eff}$'s for the subgiant stars provided best agreement with the values given
by line excitation. Adopted values for 47 Tuc are given in Table 2 of Carretta
et al. (2004).

Values of the surface gravity were derived from the location of stars in the
colour magnitude diagram; an age of 14 Gyr and corresponding masses were
assumed.

Estimates of the microturbulent velocity $v_t$ for each star were derived, as
usual, by eliminating trends of abundances with expected line strengths. Again,
average values for each group of stars in similar evolutionary phases were
adopted in NGC 6397 and NGC 6752.

Finally, the overall model metallicities [A/H] were chosen as equal to the Fe
abundances that best reproduce the measured equivalent widths ($EW$), using the
model atmospheres from the Kurucz (1995) grid with the overshooting option
switched off.

\section{ANALYSIS}

\subsection{Carbon and isotopic ratios}

Carbon abundances for the program stars were obtained from a comparison of
observed and synthetic spectra in the wavelength region from 4300~\AA\ to
4340~\AA. This spectral region includes the band head of the (0-0),(1-1) and
(2-2) bands of the A$^2$$\Delta$-X$^2$$\Pi$ transitions of CH. We used newly
derived line lists from Lucatello et al. (2003). Briefly, the starting line list was
extracted from Kurucz's database (Kurucz CD-ROM 23, 1995), including
atomic species and molecular lines of C$_2$, CN, CH, NH and OH. A few lines,
missing in the Kurucz database, were added from the solar tables (Moore,
Minnaert \& Houtgast 1966); when unidentified, we arbitrarily attributed these
lines to Fe I, with an excitation potential of EP=3.5 eV.

The  dissociation potential of CH has been determined with high accuracy at
$D^0_0 = 3.464$ eV (Brzozowoski et al. 1976), while band oscillator strenghts
were modified in order to reproduce the observed solar spectrum (Kurucz et al.
1984), using the solar carbon abundance of Anders and Grevesse (1989).  We
found that, in order to have a good match, a corrective factor of $-$0.3 dex in
the $\log gf$ values of the electronic transitions and a shift of $-$0.05~\AA\ in
wavelength were required, with respect to the values given by Kurucz. As found
by many authors (see e.g. Grevesse \& Sauval 1998), high excitation CH lines
listed by Kurucz are missing in the spectra of the Sun and other stars, due to
pre-dissociation. We omitted from our line list those lines rising from levels 
with excitation potential over 1.5 eV.

\begin{figure} 
\psfig{figure=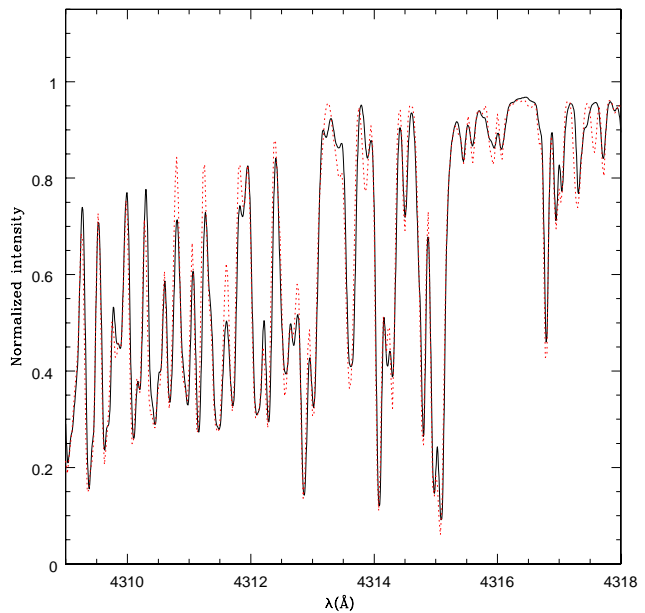,width=8.8cm,clip=}
\caption[]{The observed solar spectrum (Kurucz et al. 1984), shown as a
continuous line, with overimposed the synthetic spectrum obtained using the
adopted line list and the solar model (dotted line), in the G-band region at
$\sim$ 4309-18~\AA. Notice that the spectral region shown in this Figure
includes less than 1/4 of that used in our comparison with synthetic spectra.}
\label{f:solargband} 
\end{figure}

The excellent match of the synthetic spectrum with the observed solar spectrum
in part of the G-band region is shown in Figure~\ref{f:solargband}.

Using the appropriate atmospheric parameters, synthetic spectra in the spectral
region 4300-4340~\AA\ were computed varying [C/Fe] in steps of 0.2 dex, in the
range from 0.3 to $-$0.7 dex. A constant oxygen abundance [O/Fe] = 0 was adopted
in all these computations. The exact values affect only negligibly the
derived C abundances, since for stars warmer than $\sim 4500$ K  the coupling
of C and O is not relevant.

After the synthesis computations, the generated spectra were convolved with
Gaussians of appropriate FWHM to match the broadening mechanisms (in
particular that due to the instrumental response) of the observed spectra.
Carbon abundances were then derived from a set of 15-17 CH features within the
region under scrutiny, inspecting by eye all features, and computing an
average value for each star.

\begin{figure} 
\psfig{figure=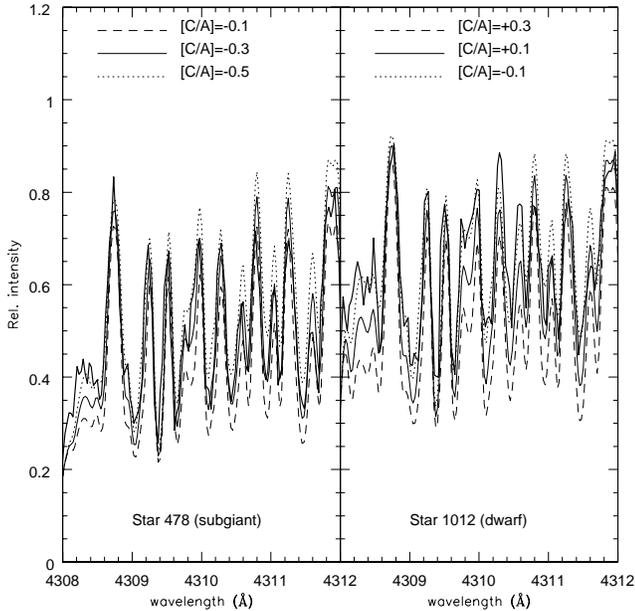,width=8.8cm,clip=}
\caption[]{Left panel: spectrum synthesis of some features of CH band in a 
subgiant star of 47 Tuc. The heavy solid line is the observed spectrum, while
dashed, solid and dotted lines are the synthetic spectra computed for three
values of the C abundances (listed on top of figure). Right panel: the same, for
a dwarf star of 47 Tuc. Note that the synthetic spectra are now computed with
different C values. All synthetic spectra were convolved with a Gaussian
to take into account the instrumental profile of observed spectra.}
\label{f:figch} 
\end{figure}

Results are summarized in Table~\ref{t:cno}, while Figure~\ref{f:figch} shows
two examples of the synthetic spectrum fits to the observed CH features for a
subgiant (star 478) and a turn-off star (star 1012) in 47 Tuc. The average
$rms$ deviations of the abundances from the individual observed features in
[C/Fe] are 0.10-0.12 dex.

Isotopic ratios $^{12}$C/$^{13}$C were estimated from spectrum synthesis in
the two regions 4228-4240~\AA\ and 4360-4372~\AA\ containing various clean
features of $^{13}$CH (see e.g. Sneden, Pilachowski and Vandenberg 1986;
Gratton et al. 2000). Synthetic spectra were computed using the C abundances
for each star derived from the G-band synthesis and the appropriate
atmospheric parameters, and a range for the $^{12}$C/$^{13}$C values. The
adopted isotopic ratios were derived as the averages from several features in 
both regions.

\subsection{Nitrogen}

\begin{figure} 
\psfig{figure=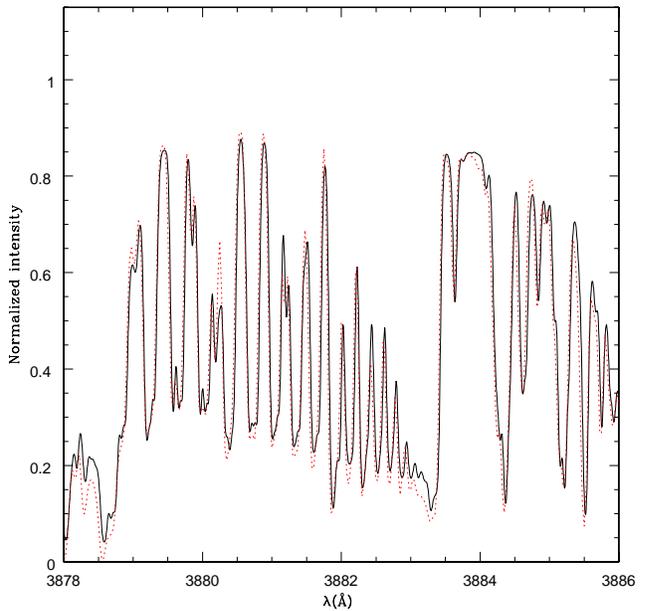,width=8.8cm,clip=}
\caption[]{The observed solar spectrum (Kurucz et al. 1984), shown as a
continuous line, with overimposed the synthetic spectrum obtained using the
adopted line list and the solar model (dotted line), in the region at
$\sim$ 3880~\AA\ including the $\Delta v=0$ bandheads of UV CN transition.}
\label{f:solarcn} 
\end{figure}

Since the violet CN band at 4200~\AA\ is vanishingly weak in warm metal-poor
stars (see Cannon et al. 1998, but see below the case of 47 Tuc), we used the
UV CN band and derived N abundances from a number of CN features in the
wavelength range 3876-3890~\AA, where the bandhead of the $\Delta v=0$ bands of
the UV system lies, again  using line lists optimized by Lucatello et al. (2003).
These lists use a CN dissociation potential of 7.66 eV from Engleman and Rouse
(1975); a corrective factor of -0.3 dex in the $\log gf$ of the electronic
transitions was applied also in this case, with respect to the values listed by
Kurucz. In Figure~\ref{f:solarcn} the comparison between the observed solar
spectrum (Kurucz et al. 1984) and  the synthetic spectrum computed with the
optimized line list is shown.
 
Carbon abundances derived above from synthesis of the CH bands were adopted in
the computation of synthetic spectra, relevant for individual stars, together
with the appropriate atmospheric parameters (from Gratton et al. 2001 and
Carretta et al. 2004).

In the case of 47 Tuc, which is about 1.3 dex and 0.7 dex more metal-rich than
NGC 6397 and NGC 6752, respectively, we were able to use also the violet CN
band strengths at $\sim 4215$~\AA\ in order to estimate the N abundances, at
least in the subgiant stars. A procedure similar to that described above was
used to compute synthetic spectra in the region from 4202~\AA\ to 4226~\AA,
with the proper C abundance for each star. The N abundances resulted to be in
very good agreement with those derived from the synthesis of the 3880~\AA\
region, so for the subgiants in 47 Tuc the [N/Fe] values are those obtained as
the average of N abundances in the two regions.

No observations for the NH band were available for our program stars, apart from
the very first run (June 2000), where the setup covered the region from 3376
to 3560~\AA, missed in the following observing runs. In this run stars in both
NGC 6397 and NGC 6752, but not in 47 Tuc were observed. This choice of the
setup was driven by the consideration that the expected $S/N$ of the spectra
of the (fainter) stars in NGC 6752 and 47 Tuc was so low that likely no
meaningful abundances could be obtained.

For these spectra of NGC 6397 stars, we then prepared a line list in the
spectral range 3400-3410~\AA, where some NH lines lie, using again the solar
spectrum as a starting point; however, in order to obtain a good match, the
$\log gf$ values of NH lines in Kurucz's list had to be lowered by about 0.5 dex.
Results of the NH synthesis in stars of NGC 6397 are given in the next Sect.

\subsection{Oxygen and Sodium}

Oxygen abundances in these warm stars were derived almost exclusively from the
permitted near-IR triplet at 7771-75~\AA, as discussed at length in Gratton et
al. (2001) and Carretta et al. (2004). Only for one subgiant in 47 Tuc 
could we measure the forbidden [O I] lines. For the other stars, the
very weak [O I] line was masked by much stronger telluric features, so that no
reliable abundance could be derived. Final abundances and upper limits are
given in Table~\ref{t:cno}, corrected for non-LTE effects as described in
Gratton et al. (1999), from statistical equilibrium calculations based on
empirically calibrated collisional H I cross sections. The appropriate
corrections were also applied to the Na abundances, derived from the strong
doublet at 8183-94~\AA.

\section{RESULTS}

Derived abundances for C, N, O and isotopic ratios  $^{12}$C/$^{13}$C for stars
in NGC 6397, NGC 6752 and 47 Tuc are listed in Table~\ref{t:cno}. 
Carbon isotopic ratios could not be reliably
derived for stars in NGC 6397 and dwarfs in NGC 6752; only upper limits for C
and N abundances were obtained for dwarf stars in NGC 6397, due to the weakness
of the features and the low $S/N$ ratio in the blue region of the spectra.

Table~\ref{t:cno} also lists, for an easier comparison of the relevant elements
involved in H-burning at high temperatures, the abundances of  Na taken
from the previous  papers of this series (Gratton et al. 2001, Carretta et al.
2004).  For the 6 subgiants in NGC 6397 observed in July 2002, newly derived
Na and O abundances are also shown in this Table, where stars are ordered
according to increasing Na abundances.

{\tiny
\begin{table*}
\caption[]{Abundances of C, N, O, Na, and isotopic ratios $^{12}$C/$^{13}$C 
in stars of 47 Tuc, NGC 6752 and NGC 6397 }
%\begin{tabular}{rccrrlrclrcrr}
\begin{tabular}{rccrcr}
\hline
Star   & [C/Fe]  & [N/Fe]& [O/Fe] & [Na/Fe]&   $^{12}$C/$^{13}$C \\

\hline
\\

\multicolumn{6}{c}{NGC 6397 - dwarfs}      \\
\\

  1905 &$<$+0.50& $<$2.0&   +0.24 &  +0.09 &      \\ 
202765 &$<$+0.50& $<$1.5&   +0.33 &  +0.13 &      \\ 
201432 &$<$+0.50& $<$1.5&   +0.21 &  +0.15 &      \\ 
  1543 &$<$+0.50& $<$1.5&   +0.28 &  +0.28 &      \\ 
  1622 &$<$+0.50& $<$2.0&   +0.23 &  +0.35 &      \\ 

\\

\multicolumn{6}{c}{NGC 6397 - subgiants}      \\
\\
   706 &   +0.10& $-$0.5&   +0.54 &$-$0.48 &      \\
   760 &   +0.10& $-$0.5&   +0.39 &$-$0.43 &      \\
   777 &   +0.15&   +0.2&   +0.39 &$-$0.25 &      \\
   737 &   +0.15&   +1.4& $<$0.06 &  +0.19 &      \\
   793 & $-$0.10&   +1.2& $<$0.06 &  +0.23 &      \\
   703 &   +0.00&   +1.5& $<$0.06 &  +0.30 &      \\
206810 & $-$0.07&   +1.3& $<$0.31 &  +0.32 &      \\
   729 & $-$0.10&   +1.4& $<$0.06 &  +0.48 &      \\
   669 &   +0.01&   +1.3&   +0.31 &  +0.53 &      \\

\\
\multicolumn{6}{c}{NGC 6752 - dwarfs}      \\
\\

  4428 &   +0.09&    +1.1&   +0.33 &$-$0.29 &     \\ 
  4383 &   +0.12&    +1.2&   +0.57 &$-$0.18 &     \\ 
202316 &   +0.12&    +1.5&   +0.27 &$-$0.06 &     \\ 
  4341 &   +0.21&    +1.4&   +0.20 &  +0.20 &     \\ 
  4661 & $-$0.20&    +1.4& $-$0.35 &  +0.29 &     \\ 
  4458 & $-$0.20&    +1.5&   +0.02 &  +0.31 &     \\ 
  5048 & $-$0.20&    +1.5& $-$0.30 &  +0.37 &     \\ 
  4907 & $-$0.20&    +1.5& $-$0.25 &  +0.58 &     \\ 
200613 & $-$0.20&    +1.7&         &  +0.62 &     \\ 

\\

\multicolumn{6}{c}{NGC 6752 - subgiants}      \\
\\

  1406 & $-$0.13&    +0.0&   +0.44 &  +0.02 &  9  \\
  1665 & $-$0.28&    +1.0&         &  +0.10 & 11  \\
  1445 & $-$0.42&    +1.2&         &  +0.14 &  5  \\
  1400 & $-$0.25&    +1.0&   +0.38 &  +0.20 &  9  \\
  1563 & $-$0.33&    +1.3&   +0.42 &  +0.26 &  5  \\
  1461 & $-$0.32&    +1.3&         &  +0.28 &  5  \\
202063 & $-$0.37&    +1.2&   +0.53 &  +0.31 &  3  \\
  1460 & $-$0.49&    +1.4& $<$0.29 &  +0.42 &  5  \\
  1481 & $-$0.51&    +1.3&         &  +0.51 &  4  \\

\\

\multicolumn{6}{c}{NGC 104 - dwarfs}         \\
\\

1081   &$-$0.13 &$-$0.50&    +0.57& $-$0.34& $>$10\\
1012   &$-$0.10 &$-$0.30&    +0.48& $-$0.14& $>$10\\
 975   &$-$0.13 &$-$0.30&    +0.40&   +0.22& $>$10\\

\\

\multicolumn{6}{c}{NGC 104 - subgiants}      \\

\\

   482 &$-$0.16 &  +0.10&    +0.61&  +0.06 &   12 \\
206415 &$-$0.11 &$-$0.25&    +0.52&  +0.10 &   10 \\
201075 &$-$0.12 &$-$0.30&    +0.41&  +0.11 &    9 \\
   433 &$-$0.32 &  +1.10&	  &  +0.24 &   10 \\
   456 &$-$0.28 &  +1.00& $<+0.19$&  +0.28 &   12 \\
201600 &$-$0.50 &  +1.10& $<+0.09$&  +0.30 &   10 \\
   435 &$-$0.31 &  +0.70& $<-0.19$&  +0.31 &    9 \\
   429 &$-$0.35 &  +0.50&  $-$0.01&  +0.31 &    6 \\
   478 &$-$0.30 &  +0.90&	  &  +0.37 &    9 \\
\hline
\end{tabular}
\begin{list}{}{}
\item[] Values of [N/Fe] in the subgiants of 47 Tuc are the average of the
abundances estimated from the synthesis of the 3883~\AA\ and the 4215~\AA\ 
regions.
\end{list}                                     
\label{t:cno}
\end{table*}
}

In the following, some features of the analysis of individual clusters are
discussed.

\paragraph{47 Tuc}

Being much more metal-rich than the other two clusters, 47 Tuc is the only one
for which we were able to obtain meaningful lower limits of the isotopic
ratios $^{12}$C/$^{13}$C for dwarf stars. The values found are listed in
last column of Table~\ref{t:cno}.

For the three dwarfs, the rather low quality of the spectra in the blue
hampered a precise determination of the isotopic ratio. Hence, we choose to
smooth somewhat the spectra, degrading the resolution to enhance the $S/N$.
However, even in this case, the best result we could secure is that the ratio
$^{12}$C/$^{13}$C is $>10$ in these turn-off stars.

It should be noticed that these are, to our knowledge, {\it the first
determinations}, of the $^{12}$C/$^{13}$C isotopic ratios in stars less
evolved than the RGB-bump in GCs.

\paragraph{NGC 6752}

For dwarfs in NGC 6752, the quality of spectra does not allow a clearcut
determination of the C abundances in the O-poor dwarfs, which are very rich in
N but with low C abundances. Hence, in these warm stars, the features of CH are
rather weak, and we adopted the following procedure, in order to obtain a more
reliable estimate.

The spectra of individual dwarf stars with low or not detected oxygen were
summed up and this coadded spectrum was then used to derive an abundance of C
through comparison with synthetic spectra. Our best estimate is [C/Fe]$=-0.2
\pm 0.1$. Analogously, the N abundances were then derived from the region
3876-3890~\AA\ using [C/Fe]$=-0.2$ dex and synthetic spectra computed with
different values of the [N/Fe] ratio. Notice that apparently there are no
N-poor dwarfs, and only one N-poor subgiant, in our sample.

\paragraph{NGC 6397}

\begin{figure} 
\psfig{figure=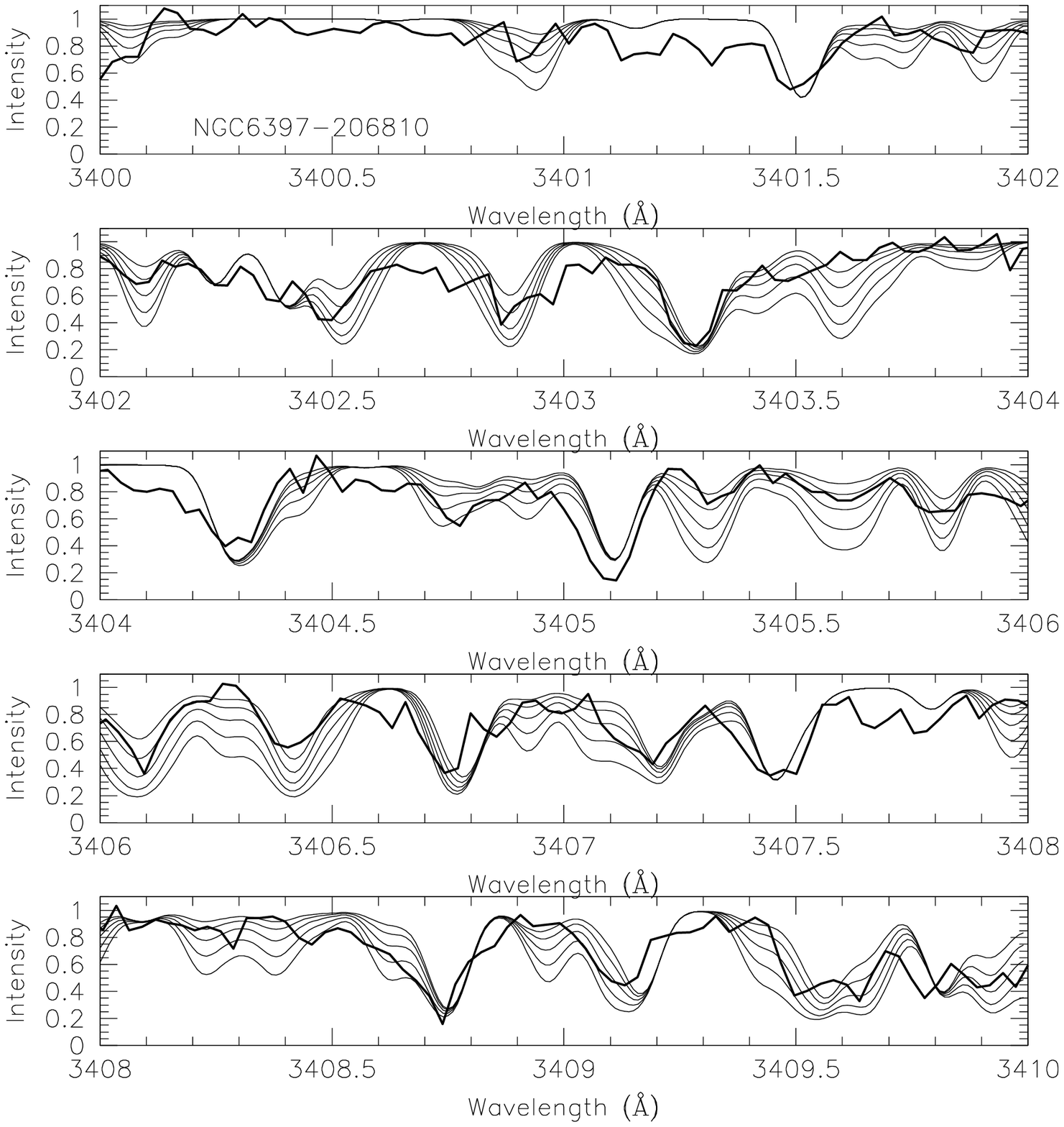,width=8.8cm,clip=}
\caption[]{The observed spectrum of the subgiant 206810 in NGC 6397 (heavy
solid line) in the spectral region 3400-3410~\AA. The thin solid lines are
synthetic spectra computed by using values of [N/Fe]= 1.0, 1.25, 1.50, 1.75 and
2.0 from top to bottom, respectively. The NH lines are clearly observed.} 
\label{f:nh1} 
\end{figure}

In some stars of NGC 6397, acquired with the bluest setup during our first run,
we were able to investigate the NH molecular bands. The great advantage of
using the hydrides bands in metal-poor clusters like NGC 6397 is that bands of
bi-metallic  molecules like  CN become vanishingly weak at low metallicity,
due to their quadratic dependence on the metal abundance. Figure~\ref{f:nh1}
shows the observed spectrum of the subgiant star 206810 as compared to five
synthetic spectra computed with different [N/Fe] ratios. The lines of NH are
clearly observed and the best match is obtained with the synthetic spectrum
computed with [N/Fe]$\simeq 1.3$ dex, in very good agreement with the value
that was derived from the CN bands. This supports our derivation of the N
abundance for the subgiants in this cluster.

\begin{figure} 
\psfig{figure=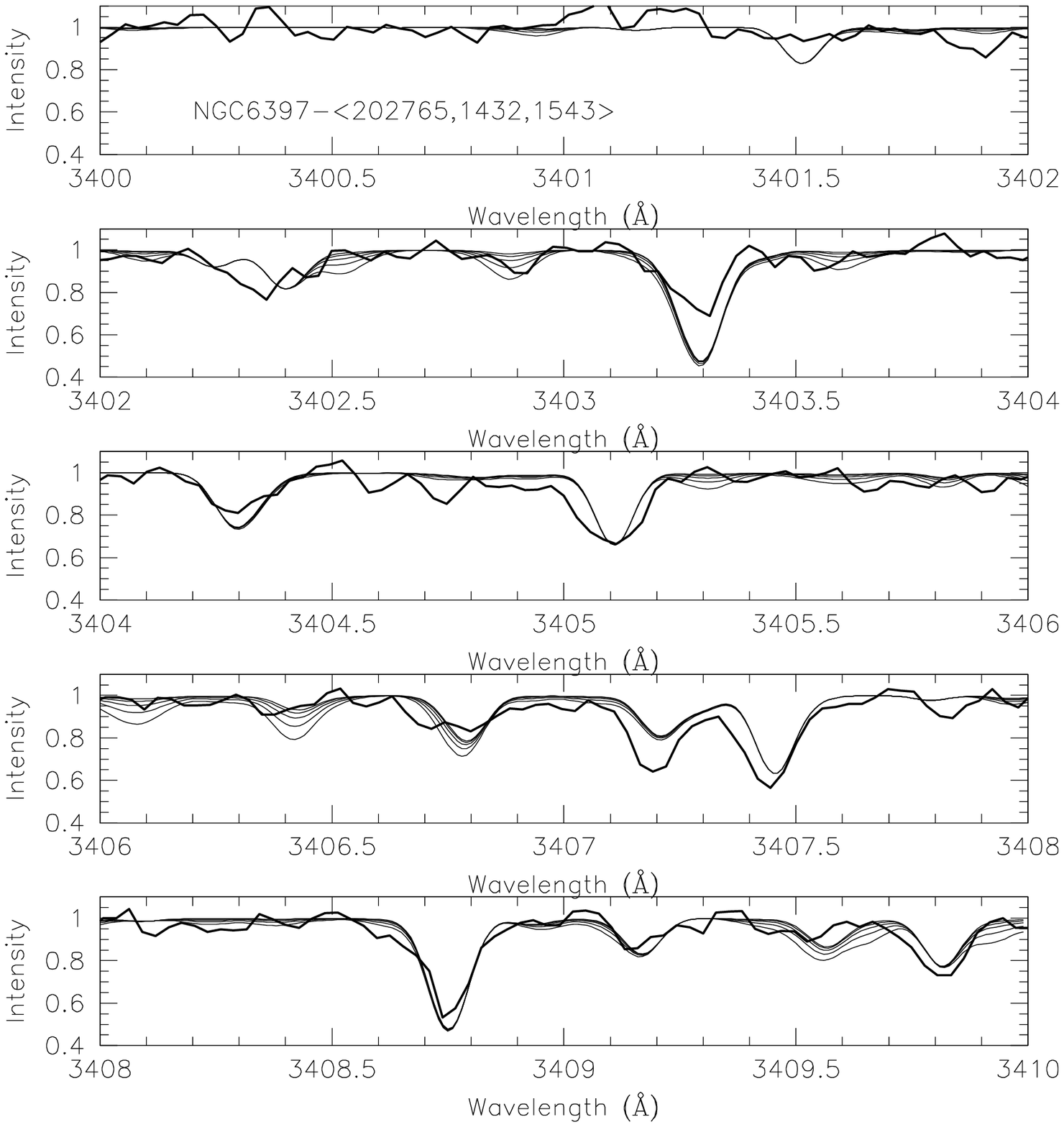,width=8.8cm,clip=}
\caption[]{The same as in previous Fig. but for the average spectrum of 3
dwarfs in NGC 6397, namely stars 202765, 201432 and 1543.}
\label{f:nh2} 
\end{figure}

In Figure~\ref{f:nh2} the same comparison is made with the average spectrum
obtained from the three dwarfs in NGC 6397 having the best spectra in the UV,
namely stars 202765, 201432 and 1543. The resulting average spectrum was
decontaminated for a relevant (about 20\% of the total value) contribution of
scattered light due to the sky, not properly taken into account by our spectrum
extraction procedure (note that these lines lie at the extreme UV edge of the
observed spectrum).

In this average spectrum we cannot firmly conclude that NH lines are actually
observed: only some lines, but not all, are detected, and even these are very
close to the noise level. The comparison with synthetic spectra in this region
shows that a reasonable fit might be achieved at [N/Fe]$\leq 1.5$ dex: a value
greater than 2.0 dex is clearly excluded. This upper limit is however more
stringent that that derived from the CN lines.

\section{DISCUSSION}

\subsection{The light elements: C and N }

\begin{figure} 
\psfig{figure=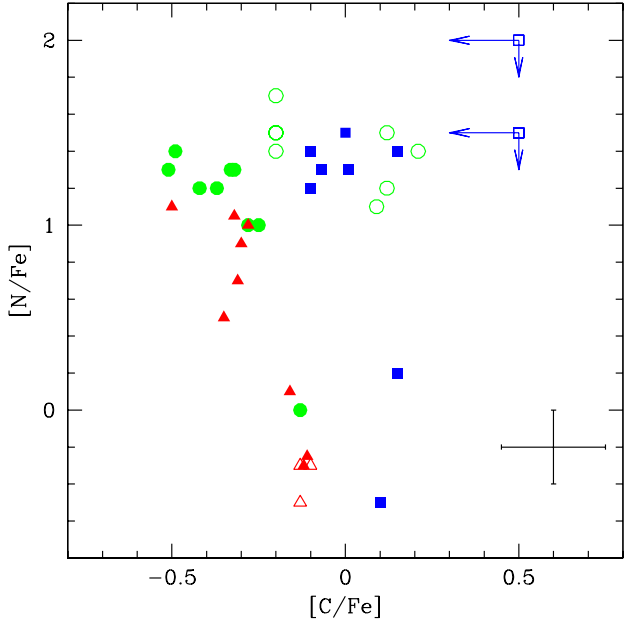,width=8.8cm,clip=}
\caption[]{Run of the [C/Fe] ratio as a function of [N/Fe] for stars in 47 Tuc
(red triangles), NGC 6752 (green circles) and NGC 6397 (blue squares). Open
symbols represent dwarfs and filled symbols are subgiant stars, for all three
clusters. Arrows represent upper limits in N, C abundances. Typical error bars
are also shown.}
\label{f:chcn} 
\end{figure}

Variations of the abundances of C and N in the examined stars, as derived from
the features of the G-band and UV CN band, respectively, are shown in
Figure~\ref{f:chcn}. Typical error bars are also shown; they are conservative
estimates, including both the scatter from the observed individual lines of CH
and CN and the effect (almost negligible) of errors in the adopted atmospheric
parameters (see Gratton et al. 2001 and Carretta et al. 2004 for the estimates
of these uncertainties).

In 47 Tuc the dwarfs are clustered around [C/Fe]$\sim -0.15$ dex, [N/Fe]$\sim
-0.4$ dex, while the subgiants seem to be divided into two distinct groups, one
with low N-high C and the other with low C-high N. Abundances of N and C seem
to be anticorrelated in the other two clusters tt, even if for NGC 6397 we
derived only upper limits for dwarfs, and the evidence of anticorrelation is
somewhat weaker in the subgiants, with respect to stars in NGC 6752 and 47
Tuc\footnote{Note that at [N/Fe]$=-0.5$ and [C/Fe]$=+0.10$ there are $two$
subgiant stars in NGC 6397, where in Figure~\ref{f:chcn} only one point is
displayed.}.

Even taking into account the small number statistics, it does not seem 
premature to
conclude that in all 3 clusters there are a few subgiants with very low N
abundances, well separated from high-N/low-C subgiants. The average C
abundance of the 3 low-N subgiants in 47 Tuc is [C/Fe]$=-0.13$ dex ($rms=0.03$
dex, 3 stars), whereas the 6 subgiants with high N abundances have an average
of [C/Fe]$=-0.34$ ($rms$=0.08). This difference in [C/Fe] is very similar to
the one between the N-poor subgiant in NGC 6752 ([C/Fe]$=-0.13$) and the
average obtained from the other (N-rich) subgiants: [C/Fe]$=-0.37$
($rms=0.09$, 8 stars). The spread in C abundances is smaller (about 0.15 dex)
for subgiants in NGC 6397: in this case we obtain [C/Fe]$=+0.12$ dex
($rms=$0.12 dex, 3 stars) and [C/Fe]$=-0.02$ dex ($rms=$0.10 dex, 6 stars)
respectively for N-poor and N-rich stars.

From these numbers we note what is immediately apparent in Figure~\ref{f:chcn}:
while the spread in [N/Fe] is well above 1 dex, in each cluster there is a
rather small variation in C abundances. Since the C/N ratio in C-rich, N-poor
stars is roughly solar ($\sim 0.6$ dex), N in N-rich stars cannot be produced
only by transformation of C into N. Furthermore, even if carbon is a minority
species in these stars, the residual C observed in N-rich stars is much more
than that expected for material processed by CN-cycle at high temperature
([C/Fe]$\lsim -0.8$; see Langer et al. 1993)

\begin{figure} 
\psfig{figure=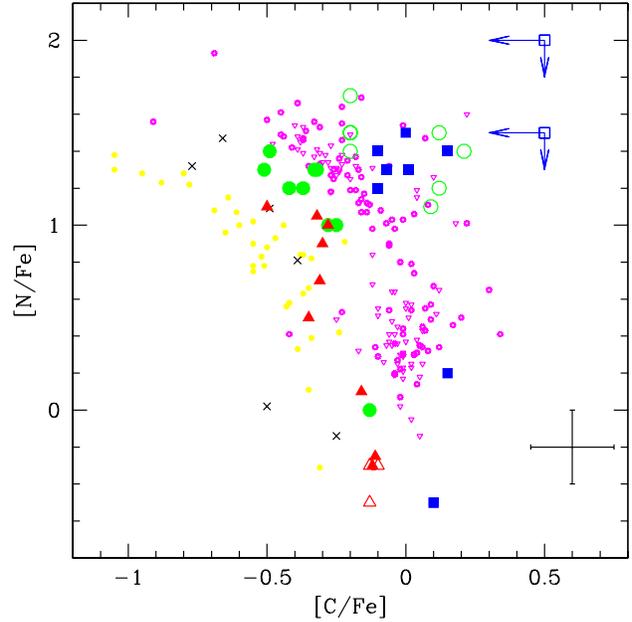,width=8.8cm,clip=}
\caption[]{Run of the [C/Fe] ratio as a function of [N/Fe]. For our stars in
47 Tuc, NGC 6397 and NGC 6752 symbols are as in Fig. 6, with typical error
bars shown. For literature data (all smaller symbols), filled yellow circles
are SGB stars in M 5 from Cohen, Briley and Stetson (2002), black crosses are
main sequence turn-off stars in M 13 from Briley et al. (2004a), magenta empty
triangles are M 71 turn-off stars from Briley and Cohen (2001) and magenta
empty exploded stars are 47 Tuc MS stars from Briley et al. (2004b).}
\label{f:lowr} 
\end{figure}  

\subsection{Comparison with previous works}

The anti-correlation between C and N abundances, already known from low
resolution spectra, is confirmed by our high dispersion spectra. To show this,
we reproduced in Figure~\ref{f:lowr} a similar plot shown by Briley et al.
(2004a) with abundances of C and N for unevolved or slightly evolved stars in a
number of clusters. While zero-point offsets are likely present between our
data set and the [C/Fe] and [N/Fe] values derived by Briley et al. (as shown
by mean ridge lines for 47 Tuc), the behaviour is essentially the same. In all
clusters examined so far, variations in C and N are anti-correlated with each
other, with N showing large spreads, with respect to the more modest scatter
in C abundances. Only among the SGB stars in M5 studied by Cohen et al. (2002)
does the spread in C seem to equal the spread in N, and the most C-poor stars have
a C depletion close to that expected by complete CN-cycling.

Is this C-N anticorrelation tied to evolutionary phenomena occurring within the
stars themselves or are we seeing the outcome of an already established
nucleosynthesis implanted early in the material? To answer this question, we
have to look into the evolutionary status of our program stars and seek for
relationships with luminosity.

\begin{figure} 
\psfig{figure=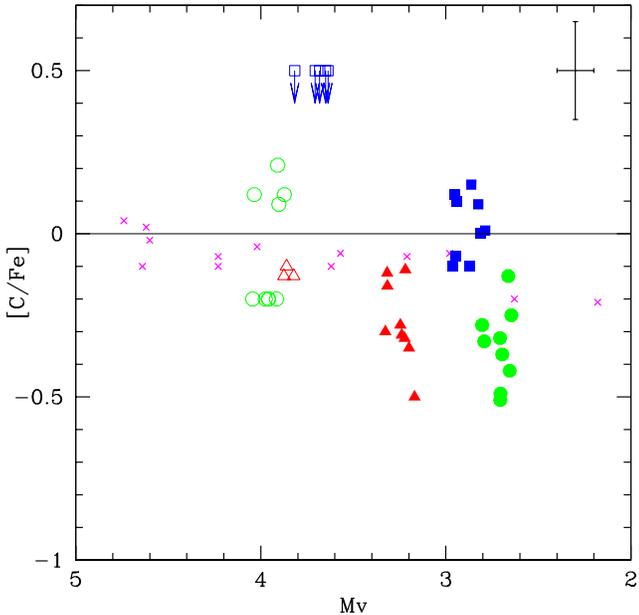,width=8.8cm,clip=}
\caption[]{Run of the [C/Fe] ratio as a function of absolute magnitude for
program stars in NGC 6397, NGC 6752 and 47 Tuc. Symbols are as in Fig. 6 .
Small (magenta) crosses are the field stars of the Gratton et al. (2000)
sample with metallicity in the range $-2.0 \leq$ [Fe/H] $\leq -1$.}
\label{f:fieldcmv1} 
\end{figure}  

In Figure~\ref{f:fieldcmv1} we plotted the [C/Fe] values of all program stars
in the three clusters as a function of the absolute magnitude of the stars.
Distance moduli are those corrected for the effect of binarity in Gratton et
al. (2003). As a reference, we also plotted halo field stars from Gratton et
al. (2000), in the same luminosity range; we only considered field stars in the
range $-2.0 \leq$ [Fe/H] $\leq -1$, that closely matches the metallicity range
of our three globular clusters. This figure shows that the C abundance drops moderately
(less than a factor of 2) both in field and cluster stars at the expected
luminosity for the first dredge-up, in agreement with the standard stellar
evolution models.

\begin{figure} 
\psfig{figure=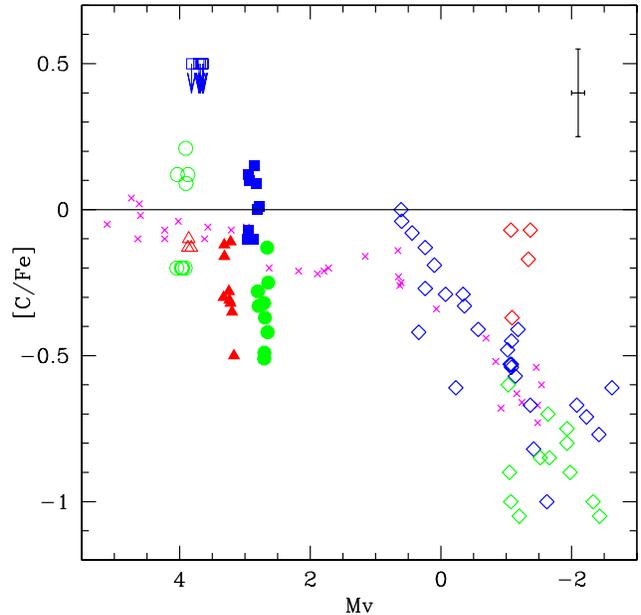,width=8.8cm,clip=}
\caption[]{Run of the [C/Fe] ratio as a function of absolute magnitude for
program stars in NGC 6397, NGC 6752 and 47 Tuc. Symbols are as in Fig. 6. 
Small (magenta) crosses are the field stars of the Gratton et al. (2000)
sample with metallicity in the range $-2.0 \leq$ [Fe/H] $\leq -1$. Open (blue)
diamonds are the bright giants in NGC 6397 from Briley et al. (1990); open
(green) diamonds are red giants in NGC 6752 from Suntzeff and Smith (1991);
open (red) diamonds are red giants in 47 Tuc from Brown et al. (1990).}
\label{f:fieldcmv2} 
\end{figure}

Admittedly, the magnitude range is rather limited, but we can extend it by
using literature data available for bright giants in the studied clusters. In
Figure~\ref{f:fieldcmv2} we added to our data also C measurements for red
giants in NGC 6397 (Briley et al. 1990), NGC 6752 (Suntzeff and Smith 1991)
and 47 Tuc (Brown et al. 1990). In all cases, the absolute magnitude scale is
that defined by Gratton et al. (2003).

Despite the heterogeneity of data sources and methods used to obtain the C
abundances (low-dispersion spectroscopic measurements of the G-band in stars of
NGC 6397 by Briley et al. 1990; infrared spectra of the first overtone CO bands
for stars studied by Suntzeff and Smith 1991 in NGC 6752; moderately high
resolution spectra of the CN red system and of the G-band for stars in 47 Tuc
by Brown et al. 1990; synthesis of high resolution spectra of the UV CN system
and of the G-band in our program stars), and offsets between the bright giants
and the unevolved stars might be present, the overall pattern among cluster
stars seems to follow rather well the one defined by "undisturbed" field
stars. Namely, the second step-like decrease in [C/Fe] ratios around the red
giant branch bump ($M_V\sim 0.5$) seems to be present also for cluster stars.
This is explained (e.g. Sweigart and Mengel 1979; Charbonnel 1995; Gratton et
al. 2000) as the onset of a second mixing episode during the red giant
evolution of a population II star, once the molecular weight barrier
established by the retreating convective envelope is wiped out by the
advancing shell of H-burning. From now on, CN-processed material is able to
reach the surface layer, where a further decrease of C is visible, as shown by
Figure~\ref{f:fieldcmv2}. 

We conclude that in spite of a larger (intrinsic) scatter, the same mixing
episodes observed among field stars can be traced also among cluster stars.

This result is not new. Very recently, Smith and Martell (2003) used the same
field sample by Gratton et al. (2000) and literature data for C abundances in
red giants in M 92, NGC 6397, M 3 and M 13 to show that the rate of decline of
[C/Fe] on the RGB as a function of $M_V$ is very similar between cluster and
halo field giants. Our study, however, has the advantage of sampling regions
along the RGB well below the so called bump in the RGB luminosity function,
where standard theories for extra-mixing (see e.g. Sweigart and Mengel 1979)
fix the threshold in magnitude for the onset of additional mixing. In globular
clusters the chemical anomalies can be traced down to very faint magnitudes
and we clearly detect a steady increase in the average C abundance going from
red giants to subgiants and to dwarf stars.

Figure~\ref{f:fieldcmv2} also shows another well known feature: cluster
stars seems to reach more extreme C depletions than those experienced by field
analogs, as clearly indicated by red giants in NGC 6752 and, partly, by
giants in NGC 6397. 

\begin{figure} 
\psfig{figure=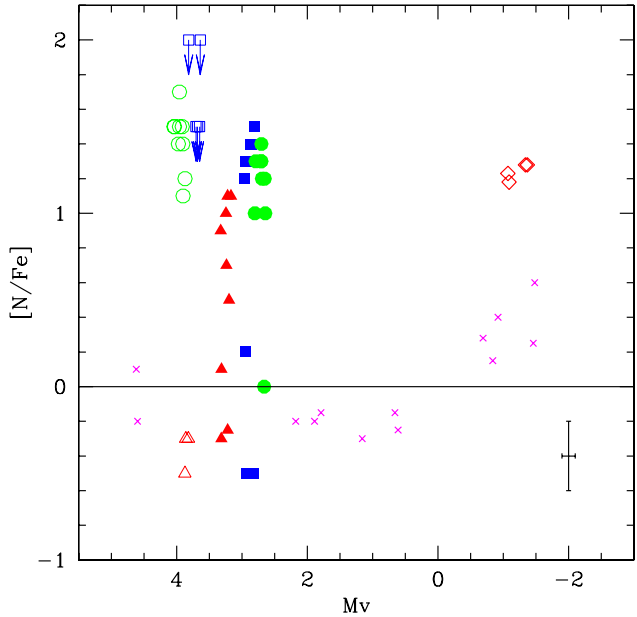,width=8.8cm,clip=}
\caption[]{Run of the [N/Fe] ratio as a function of absolute magnitude for
program stars in NGC 6397, NGC 6752 and 47 Tuc, field stars (Gratton et al.
2000) and cluster red
giants from literature (Brown et al. 1990). Symbols and color codes are as in
the previous Figure.}
\label{f:fieldnmv1} 
\end{figure}  

The analogous plot for N abundances, in Figure~\ref{f:fieldnmv1}, also reveals
an extreme behaviour of cluster stars. Spreads in [N/Fe] are very large in
cluster dwarfs and subgiants with respect to field stars of similar
evolutionary status. Moreover, when coupled with literature data (from Brown
et al. 1990), a hint for increasing [N/Fe] at increasing luminosity seems to
appear for stars in 47 Tuc. On the other hand, no clear indication of such
increase in NGC 6752 is present.

\begin{figure} 
\psfig{figure=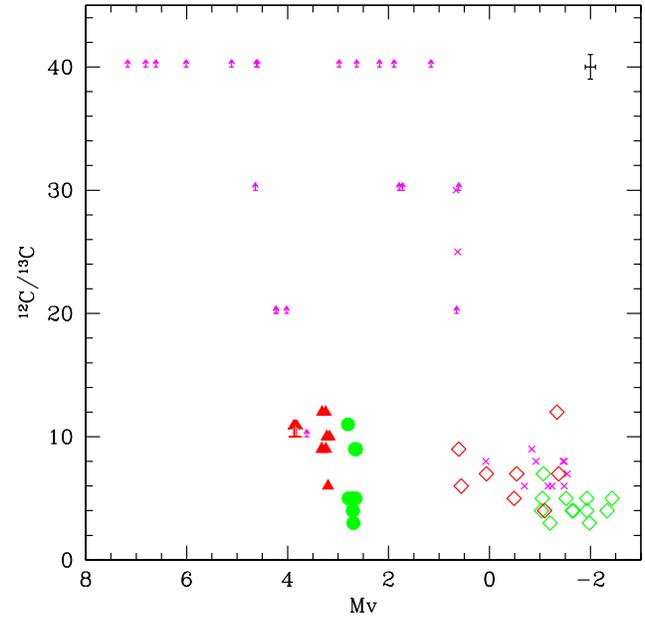,width=8.8cm,clip=}
\caption[]{Run of the isotopic ratio $^{12}$C/$^{13}$C as a function of $M_V$ for
program stars in NGC 6397, NGC 6752 and 47 Tuc. Symbols are as in the previous 
Figures. Small (magenta) crosses and upper limits are the field stars of the 
Gratton et al.
(2000) sample with metallicity in the range $-2.0 \leq$ [Fe/H] $\leq -1$. 
Open (green) diamonds are red giants in NGC 6752 from Suntzeff and Smith
(1991); open (red) diamonds are red giants in 47 Tuc from Brown et al. (1990)
and Shetrone (2003).}
\label{f:fieldccmv1} 
\end{figure}  

Finally, Figure~\ref{f:fieldccmv1} shows the isotopic ratios $^{12}$C/$^{13}$C
measured in program stars as compared to the field database of Gratton et al.
(2000), as well as the literature data available for these clusters.
Also in this case, cluster stars show extremely low values of the isotopic
ratios, when compared to the field stars of similar magnitude. However, the
interesting feature here is that the $^{12}$C/$^{13}$C values are not at the
CN-cycle equilibrium value, not even for stars that are extremely N-rich. For
these stars one would expect a value of $\sim 3$, at odds with our findings.
Even considering literature data for bright giants, their isotopic ratios seem
to be somewhat lower than those in field stars but always at a level slightly
higher than the equilibrium value.

\subsubsection{First conclusions from light elements}

In summary, the analyses of C and N and the relative abundances of the carbon
isotopes in slightly evolved globular cluster stars show that:
\begin{itemize}
\item there are moderately small variations in C abundances, anticorrelated to
(large) variations in N abundances. However, except for a few stars in M5, the C
is generally not as low as expected for CN-cycling material. In no dwarf or
early subgiant do we find C abundances as low as those observed in (all) highly
evolved RGB stars, i.e. stars brighter than the RGB-bump.
\item N shows a large spread and, with the cautionary warning of a rather
limited range in sampled magnitudes, does not seem to have a particular
increase at the first dredge-up position. Note that a large
fraction of dwarfs and early subgiants have N abundances as high as those
observed in highly evolved RGB stars (Figure~\ref{f:fieldnmv1}).
\item the isotopic ratios $^{12}$C/$^{13}$C are low but do not reach the
equilibrium value of $\sim 3$.
\end{itemize}

From these observations, it is already clear that 
no unevolved star has the same surface composition
as highly RGB stars, as required by some
recent scenarios advocated to explain the O-Na anticorrelation (Denissenkov and
Weiss 2004). Rather, the observed pattern resembles much more that predicted by
the evolutionary models of the most massive intermediate-mass,  low
metallicity AGB stars (e.g., Ventura et al. 2002, 2004). According to these
models, these stars have complete CNO cycling at the base of the convective
envelope and the coupling of nucleosynthesis plus release of processed matter
into the protocluster clouds or onto second-generation stars is able to
produce a surface composition where C is depleted, but not to values as low as
those expected from CN-cycling, because some fresh $^{12}$C produced by
triple$-\alpha$ reactions is dredged up to the surface, N is enhanced and the
carbon isotopic ratio approaches the equilibrium value, due to the large
enhancement of $^{13}$C. Apparently, this is almost exactly the chemical
composition of the unevolved cluster stars under scrutiny.

At this point of our discussion, however, this assertion is not yet proven,
because dilution with material not contaminated by CNO burning could be
invoked to explain the observed trends for C and N abundances. Observations of
heavier elements involved in high temperature $p-$capture reactions may give a
deeper insight.

\subsection{Oxygen and Sodium}

Looking now at heavier elements (O and Na), we proceed along a path of
stronger Coulomb barriers. The temperatures involved are much higher and we are
considering deeper regions in the H-burning shell.

\begin{figure} 
\psfig{figure=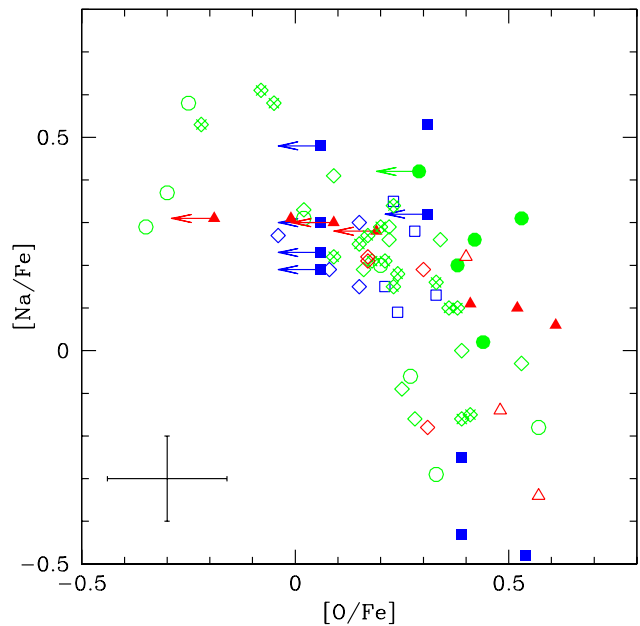,width=8.8cm,clip=}
\caption[]{Run of the [Na/Fe] ratio as a function of [O/Fe], for stars in 
47 Tuc, NGC 6752 and NGC 6397. Symbols for our program stars are as in Fig. 6.
Literature data are as follows: (green) diamonds with crosses
inside are bright red giants from the extensive study by Yong et al. (2003),
open (blue, green and red) diamonds are stars of NGC 6397, NGC 6752 and 47 Tuc,
respectively, from  Norris and Da Costa (1995), Carretta (1994) and Castilho et
al. (2000), as described in the text.}
\label{f:ona2} 
\end{figure}

The well known Na-O anticorrelation (see Gratton, Sneden and Carretta 2004 for
a recent review) is summarized in Figure~\ref{f:ona2} for our program clusters.
In this Figure we also added the available literature
data, which is mostly for bright red giant stars,
even if, apart from a few cases, no systematic studies have been
performed for these 3 clusters, often used as calibrators. Abundances of Na
and O in NGC 6397 include 2 stars studied by Norris and Da Costa (1995) and 2
stars from Castilho et al. (2000). For 47 Tuc, additional data are from Norris
and Da Costa (1995) and Carretta (1994). For NGC 6752, we added 6 stars from
Norris and Da Costa (1995), 4 stars from Carretta (1994) and the bright red
giants studied by Yong et al. (2003). Individual values of Na and O for the
bump stars in NGC 6752 analyzed by Grundahl et al. (2002) have not been published
anywhere, therefore they are not used. Whenever possible, as in the Yong et
al. sample, we started from original $\log n(O)$ and $\log n(Na)$ values,
bringing them onto a homogeneous scale by using the average [Fe/H] values and
the solar values adopted in the present study.

Note that values for our dwarfs and subgiants do include corrections for
departures from LTE. For literature data, this was possible only for stars
analyzed in Carretta (1994). However, since O abundances are usually derived in
red giants from the forbidden [O I] doublet, these corrections are negligible.
NLTE corrections for Na abundance might be of more concern in giants, depending
on what lines were used in the various analyses, but the overall appearence of
Figure~\ref{f:ona2} shows that if there are some offsets, they are rather
small.

The O-Na anticorrelation is in fact very well defined for all clusters; there
seem not to be large differences among the different clusters over 
the metallicity range sampled, nor among stars of different evolutionary status
within a given cluster. For the first time, the existence of a Na-O
anticorrelation also among stars in NGC 6397 is clearly shown. This Figure
shows among slightly evolved cluster stars the same trends that have previously
been observed among the red giant stars of several globular clusters.

The difference with respect to field stars is striking. With their highly
homogeneous sample, Gratton et al. (2000) convincingly showed that field stars
have constant Na and O abundances, not anticorrelated with each other. The
obvious implication is that in cluster stars there is something else able to
alter simultaneously the abundances of these two elements. 

\begin{figure} 
\psfig{figure=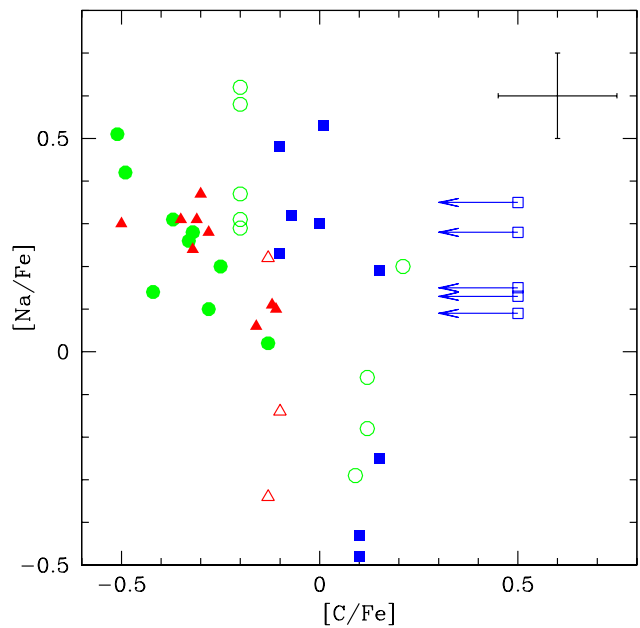,width=8.8cm,clip=}
\caption[]{Run of the [C/Fe] ratio as a function of [Na/Fe], for stars in 
47 Tuc, NGC 6397 and NGC 6752. Symbols are as in Fig. 6 .}
\label{f:cna} 
\end{figure}

\begin{figure} 
\psfig{figure=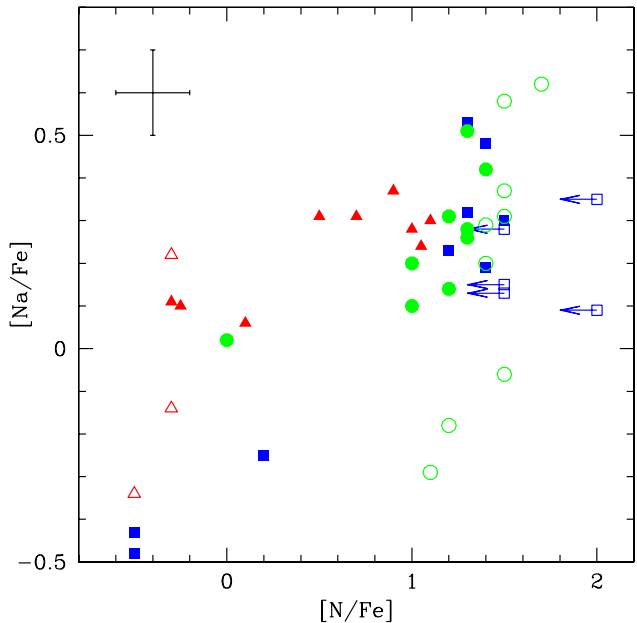,width=8.8cm,clip=}
\caption[]{Run of the [N/Fe] ratio as a function of [Na/Fe], for stars in 
47 Tuc, NGC 6752 and NGC 6397. Symbols are as in Fig. 6 .}
\label{f:nna} 
\end{figure}

To get a deeper insight, we show in Figure~\ref{f:cna} and Figure~\ref{f:nna} 
the derived abundances of C and N, respectively, for program stars as a
function of the [Na/Fe] ratio.

In all three clusters, Figure~\ref{f:cna} and Figure~\ref{f:nna} clearly show a
trend for C and N abundances to decrease and increase respectively with the
increase of Na abundances. In particular, the C-Na anticorrelation closely
mimics the well-known O-Na anticorrelation, summarized in Figure~\ref{f:ona2}.
 
\begin{figure} 
\psfig{figure=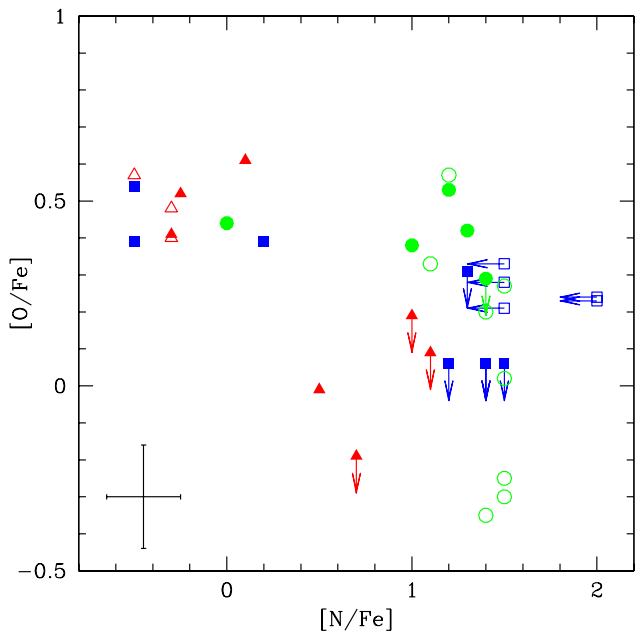,width=8.8cm,clip=}
\caption[]{Run of the [N/Fe] ratio as a function of [O/Fe], for stars in 
47 Tuc, NGC 6752 and NGC 6397. Symbols are as in Fig. 6 .}
\label{f:no} 
\end{figure}

Furthermore, while turn-off stars show a large range in Na abundances (at
almost constant C), carbon abundances are anticorrelated with Na abundances for
SGB stars. On the other hand, N abundance correlates well with sodium among
subgiant stars, even if the anticorrelation is less evident among TO stars.
Finally, O is anticorrelated with N, as shown in Figure~\ref{f:no}.

The overall distribution of light elements seems to point out that processes of
proton-capture are at work. In the atmospheres of the stars studied we are just
seeing the products of these reactions.  In this case, the line of thought is
the same as in Gratton et al. (2001): turn-off stars do not reach the
temperature regime where the ON and NeNa cycles required to produce the Na-O
anticorrelation are active, and moreover these stars have convective
envelopes that are too small to have efficiently mixed the ashes of these 
nuclear processes up to
the surface. The same conclusion holds also for subgiants.

The bottom line is that what we are seeing are the by-products of nuclear
burning and dredge-up in $other$ stars, that are now not observable, but that
have returned their elements to the intracluster medium or directly to the
surface of the presently observed stars.

\subsection{CNO arithmetic}

Having now also O abundances at hand for program stars, it is possible to test
in another way if the observed pattern of C, N, O abundances can be explained
by the CNO-cycle alone. In fact, in this hypothesis it is only the relative
content of C, N and O that may change, as a consequence of different reaction
rates; their sum has to be constant.

\begin{figure} 
\psfig{figure=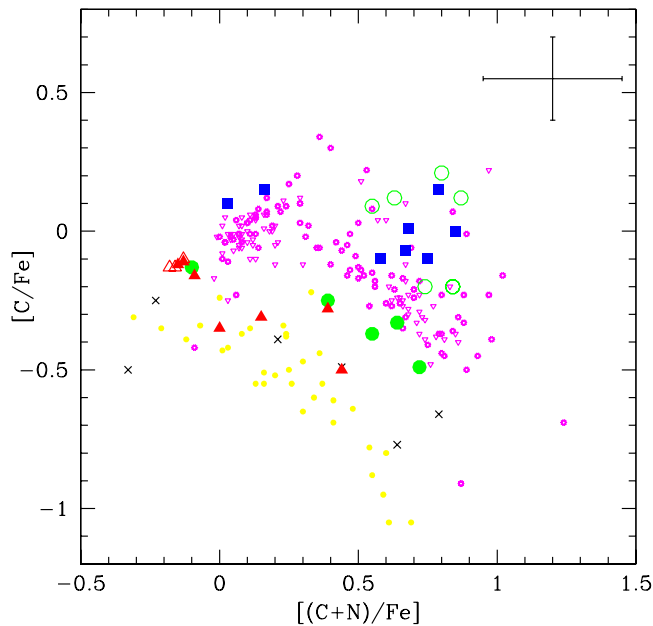,width=8.8cm,clip=}
\caption[]{Run of the [C/Fe] ratio as a function of [(C+N)/Fe]. For our stars
in 47 Tuc, NGC 6397 and NGC 6752 symbols are as in Fig. 6. For literature data
(all smaller symbols), filled yellow circles are SGB stars in M 5 from Cohen
et al. (2002), black crosses are main sequence turn-off stars in M 13 from
Briley et al. (2004a), magenta empty triangles are M 71 turn-off stars from
Briley and Cohen (2001) and magenta empty exploded stars are 47 Tuc MS stars
from Briley et al. (2004b).}
\label{f:lowr3} 
\end{figure}

In Figure~\ref{f:lowr3} the C abundances are plotted as a function of the sum
C$+$N for our program stars and, as a reference, for stars with abundances from
low dispersion studies, as in Figure~\ref{f:lowr}.

From this Fig. one has the impression that the sum C$+$N increases with
decreasing C abundance; this is confirmed by computing Kendall's $\tau$, which
implies that this anti-correlation is highly significative, at the 99.3\%
level. Hence the observed pattern cannot be due simply to a
transformation of C into N by the incomplete CN-cycle of H-burning. 

\begin{figure} 
\psfig{figure=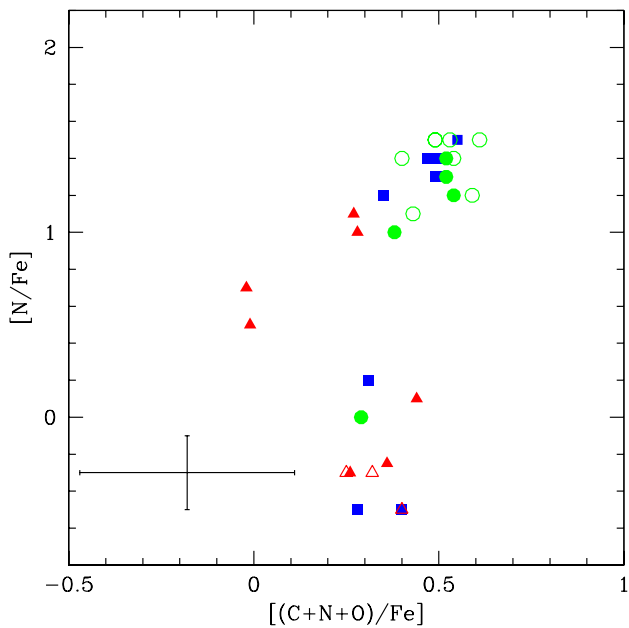,width=8.8cm,clip=}
\caption[]{Run of the [N/Fe] ratio as a function of [(C+N+O)/Fe] for our stars 
in 47 Tuc, NGC 6397 and NGC 6752; symbols are as in Fig. 6.}
\label{f:ncno} 
\end{figure}

The possible alternatives are then either that the ON-cycle is also involved,
adding N freshly produced from O, or that some variable amount of already
existing N is superimposed on the effects of C$\rightarrow$N reconversion. 
In this regard, clearcut evidence is provided by Figure~\ref{f:ncno}, where
the run of [N/Fe] as a function of the total sum C$+$N$+$O is shown.
Over a spread in N of almost 2 dex, the sum remains almost (but not exactly,
see below) constant; this by itself implies that the complete CNO-cycle 
has been at
work to produce the observed pattern. Once more, neither subgiants nor, in
particular, unevolved turn-off stars are able to forge elements (such as Na)
that require high-temperature proton-capture reactions. Moreover, they are
unable to mix in their convective envelopes such products to the atmospheric
layers; hence, we are forced to conclude that the CNO cycle was at work in
stars other than those presently observed.

The best candidates at hand are the intermediate-mass AGB stars.

Another class of possible polluters was recently suggested by Denissenkov and
Weiss (2004). According to their computations, as well as those previously
reported by Denissenkov and Herwig (2003), nucleosynthesis in IM-AGB stars
with strong O-depletion is not accompanied by large Na production (hence, the
matter is not Na-enhanced as required by the Na-O anticorrelation); instead,
strong Mg depletions are expected, and this has never been observed in
globular cluster stars. Similar results has been recently obtained by Herwig
(2004) and Fenner et al. (2004). As a way out, Denissenkov and Weiss (2004)
suggested that peculiar CNO abundances, as observed in unevolved cluster
stars, might be a result of the H-burning shell in upper RGB stars of mass
slightly larger than those presently observed in GCs, provided that they have
experienced some degree of extra-mixing (see Denissenkov and Herwig 2003),
followed by mass transfer onto less evolved stars.

\begin{figure} 
\psfig{figure=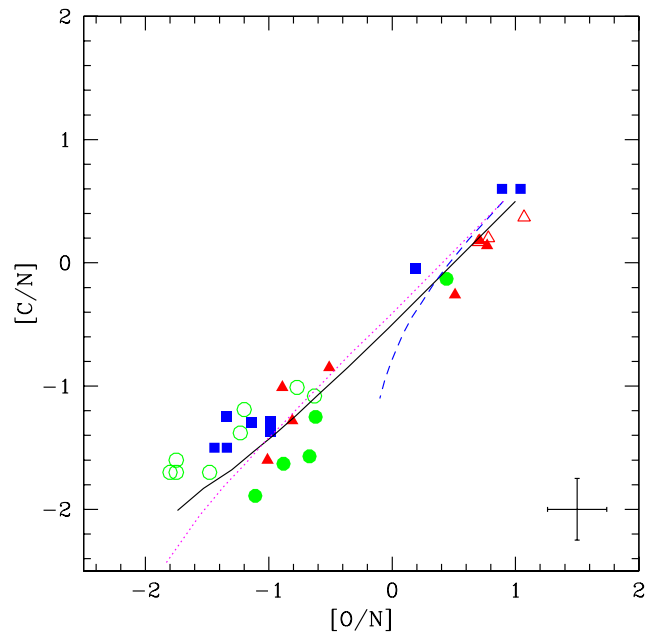,width=8.8cm,clip=}
\caption[]{Run of the [C/N] ratio as a function of [O/N] for our stars 
in 47 Tuc, NGC 6397 and NGC 6752; symbols are as in Fig. 6. Superimposed on the
data are the three models outlined in the text: a simple dilution with 
material processed through the 
complete CNO cycle (solid, black line), contamination from N-poor RGB stars
(dashed blue line) with composition typical of field RGB stars, and
contamination from N-rich upper-RGB stars experiencing very deep mixing (dotted
magenta line).}
\label{f:cnon2} 
\end{figure}  

\subsubsection{A simple dilution model}

So far, it has been shown (i) that globular cluster stars exhibit the same
mixing episodes observed for field stars; (ii) that no slightly evolved
cluster star has an abundance pattern the same as that observed among stars 
close to
tip of the red giant branch; and (iii) that there may be an excess in the sum
of C+N+O in N-rich stars, that can possibly be attributed to some $^{12}$C 
in excess with respect to the predictions of the complete CNO cycle.

To test how various schemes work to explain these observational facts, we
will consider here simple models for the dilution of the products of CNO burning at
various temperatures, and will compare their predictions with the run of the
[C/N] ratio against [O/N] ratio shown in Figure~\ref{f:cnon2}, as well as
with the other pieces of information we have gathered.

When constructing our dilution models, we started noticing that in the complete
CNO-cycle at high temperature ($40 \times 10^6$ K), at equilibrium the O
abundance is decreased much more than the C abundance (factors of about 50 and
6, respectively: Langer et al. 1993). This is quite different from the case of
the incomplete CN-cycle at low temperature ($\sim 10^6$~K), where the C
abundance is decreased by a factor of 6, as before, but the O abundance is not
modified.

\medskip
\noindent
{\bf Testing the scenario of pollution by the complete CNO-cycle}

\medskip
We started by considering the standard scenario of pollution by the complete
CNO-cycle. Let $a$ be the fraction of gas forming unevolved stars in globular
clusters that has been burnt through the complete CNO-cycle and let us assume
the remaining gas to be of primordial composition. It follows that the
abundance of C and O in the mixed gas with respect to the initial composition
is: $f= a \cdot b(C,O) + (1-a)$, where $b(C,O)$ is the abundance of C and O at
the equilibrium of the CNO-cycle (0.17 and 0.02 for C and O, respectively). The
N abundance can be derived by the constraint that the total CNO abundance is
constant.

If we assume that initially [C/N]=0.5 and [O/N]=1.0 (which is approximatively
the composition of N-poor stars in the three clusters: see
Figure~\ref{f:cnon2}), we may then predict the values of [C/N] and [O/N] for
different fractions $a$ of the gas consumed in the complete CNO-cycle, with
$a=0$ being the original composition. This trend is shown as a solid line in
Figure~\ref{f:cnon2}. This line reproduces fairly well the location of
observational points in Figure~\ref{f:cnon2}, albeit it predicts somewhat too
low C abundances for N-rich dwarfs. Dilution factors $a$\ adequate to
reproduce N-rich stars are $0.5<a<0.8$. Such values also allow reproducing
the isotopic ratio $^{12}$C/$^{13}$C$\sim 7$ observed in unevolved stars.
Moreover, with these factors we might quite easily obtain a Li abundance
roughly similar to the original one, provided that the diluting material be
Li-rich (as expected from some of the IM-AGB stars: Ventura et al. 2001).

On the other hand, with such large dilution factors, the models by Langer et
al. (1993) would predict too much Na, by more than a factor of about 10; this
is because in these models Na is produced also by $^{20}$Ne. In order to
reproduce the observations we would need that Na be forged only from $^{22}$Ne.

\medskip
\noindent
{\bf Testing the scenario of polluting RGB stars}

\medskip
Let us now use a similar model to test the scenario envisioned by Denissenkov
and Weiss (2004) of pollution by RGB stars. In this case, mixing occurs in a
fraction of upper-RGB stars and afterward a transfer of material onto the
dwarfs occurs.

Let us assume that the upper-RGB stars might have either one of the following
2 compositions: (i) a chemical composition typical of field upper-RGB stars
(N-poor, i.e. N only coming from incomplete CN-cycle) or (ii) a composition
from very deep mixing, where complete-CNO cycle and Na-enrichment are
involved. For these stars we will use the most extreme case observed (i.e.
N-rich). Note that in both groups of stars all Li is destroyed.
The starting compositions ($\log n({\rm C})/({\rm N})/({\rm O})/({\rm Na})/
({\rm C+N+O})$) assumed are then:
    8.60/8.00/8.90/6.30/9.11, 
    8.60/7.50/9.30/5.80/9.38, 
    7.50/9.33/8.40/6.80/9.38 
and 8.00/8.50/9.30/5.80/9.38 for the solar, original, N-rich and N-poor cases,
respectively.

By varying the dilution factor $a$, we obtain the abundance pattern for the two
cases original+$a \times$ (N-rich) and original+$a \times$ (N-poor). Results
are overplotted as a dotted (magenta) line and a dashed (blue) line,
respectively, over our data in Figure~\ref{f:cnon2}. The first case is very 
similar to the case made above for the complete CNO cycle, differences being 
only due to the slightly different assumptions made about the compositions.

If this scenario is correct, it would be expected that the observed points
should lie between the two lines. Actually, the line representing pollution by
N-poor stars does not reproduce the observations; on the other hand, the line
representing N-rich stars fits the data reasonably well (although not as well
the N-rich dwarfs), requiring values $0.5<a<0.8$\ similar to those obtained in
the previous subsection.

The inadequacy of models with pollution by N-poor stars is evident when
noticing that within this scheme we should expect to find C-poor, Na-poor
stars. However, these stars are not observed at all (see Figure~\ref{f:cna}).
The inference is that in globular clusters there are no dwarfs polluted by RGB
stars with a chemical composition typical of field RGB stars. Within this
scheme it should then be assumed that only stars experiencing very deep mixing
polluted unevolved stars. One would then be forced to conclude that only a
fraction of RGB stars, and only those in clusters, lose a great amount of mass,
and that these very same stars do experience also very deep mixing, likely due
to the same physical mechanism (rotation? binarity?). 

An additional problem with this scheme is that N-rich giants have generally
no Li at their surface. We would then expect that Li be depleted by a factor
of 2 to 5 in main sequence stars of globular clusters like NGC 6397, in
contrast with observations (Bonifacio et al. 2002).

There are further additional concerns in a mechanism involving pollution by RGB
stars. The lost material ends up polluting other stars. It cannot be a simple
surface pollution: in fact, in this case there should be also noticeable
differences between dwarfs and subgiants (due to different masses of the
convective envelopes) which is not observed. Since most of the unevolved stars
observed in clusters like NGC 6397 and NGC 6752 are N-rich, the total amount of
mass lost by these RGB stars should be large, about 80$\%$ of the cluster
mass. This seems unlikely, since an RGB star cannot lose more than $\sim 50\%$
of its mass, the remaining being locked in the degenerate core. Another
problem concerns the epoch when this pollution occurred. In fact, if the mass
was lost in recent times, one would expect a large numbers of young stars,
obviously not observed.
On the other hand, IM-AGB stars may eject almost 80\% of their mass
(see e.g. Marigo et al. 1998), hence the mass requirement in this case would be
met if the original initial mass function (IMF) of the cluster stars is not too
steep, allowing to form many AGB stars. Evidences for a flatter local IMF are 
discussed, in this context, by e.g. D'Antona (2004) and Briley et al. (2001).

\subsubsection{The triple-$\alpha$ scenario}

As pointed out in the previous discussion, another problem is evident from
Figure~\ref{f:cnon2}. The dilution models predict that stars having [O/N]$\sim
-1.5$\ should have [C/N]$\sim -1.9$, whereas our observations show [C/N]$\sim
-1.5$. This suggests the presence of an additional source of $^{12}$C, very
likely through the triple$-\alpha$ process. A similar excess of $^{12}$C is
also suggested by the $^{12}$C/$^{13}$C isotopic ratio. Let us in fact assume
that $c$ is the fraction of  $^{12}$C produced by triple$-\alpha$ and $(1-c)$
the fraction of $^{12}$C resulting by CNO-processing. Hence, we may write for
$^{13}$C (which is produced only by the CNO-cycle) $^{13}$C =
$\frac{(1-c)}{R_e}$, where $R_e \sim 3.5$ is the equilibrium value of the
isotopic ratio $^{12}$C/$^{13}$C. We can now re-write the fraction of $^{12}$C
from CNO by subtracting the contribution of $^{13}$C and of $^{12}$C by 
triple$-\alpha$, i.e. as 
$^{12}$C = $(1-c) -$ $^{13}$C = $1-c$ -$\frac{(1-c)}{R_e}$, hence
as $(1-\frac{1}{R_e})\times (1-c)$. 
The observed C isotopic ratio is
then $R_o$ = $(\frac{^{12}C}{^{13}C})_o = \frac{(1-\frac{1}{R_e}) \cdot(1-c) +
c}{\frac{(1-c)}{R_e}}$ from which, with simple algebra, the fraction of
$^{12}$C due to triple$-\alpha$ is $$ c = 1 - \frac{R_e}{1+R_o} .$$ By using an
observed ratio (see Table 1) of about 8, we derive that about 60\% of the C
observed is likely to come from triple$-\alpha$ burning.

This estimate compares well with what is known from models of IM-AGB stars.
In fact, typical estimates of the C abundance in NGC 6752, [C/Fe]$\sim
-0.3$\ corresponds to a mass fraction of $\sim 2\times 10^{-5}$ of
$^{12}$C, a value entirely consistent with model predictions for a 
5~M$_\odot$ star of similar metallicity (Ventura et al. 2004).

Up to now, our conclusions are based only on the observation that there is too
much C, with respect to the very large N-enhancements, and carbon
isotopic ratios too high to be explained purely with a re-arrangement of elements
involved in the CNO-cycle. However, other additional evidence comes from the O
abundances. In the low-O, low-C region of Figure~\ref{f:cnon2} the existence
of a sort of plateau also suggests that the products of triple$-\alpha$ are
involved. This is not very clear in subgiants, but quite evident in dwarfs,
whose C abundances have not been modified by the first dredge-up.

\section{CONCLUSIONS}

Summarizing, we can consider various mass ranges of likely polluters that
contributed to the chemical composition in stars presently observed:
\begin{itemize}
\item metal-poor stars in the intermediate mass range, say $1.2 \leq M \leq
3-5 M_\odot$, are the classical donors considered for CH-stars (McClure and
Woodsworth 1990). This is likely not the correct mass range for typical stars
polluting globular clusters, since they also produce consistent amounts of
$s-$process elements, that are not seen to vary in cluster stars.
\item stars less massive than this range (the range considered by Denissenkov
and Weiss 2004) are also unlikely. Apart from the possible excess of $^{12}$C
discussed above, either (i) they release a large mass to be successively
accreted, and this is unlikely, since the polluted stars would have then a
large mass range, the cluster TO would be smeared out and we would end up with
a cluster mostly composed of blue stragglers\footnote{Actually, we think that
the mechanism proposed by Denissenkov and Weiss (2004) is indeed active in a
minority of globular cluster stars belonging to binary systems, producing the
blue straggler stars with the classical McCrea (1964) mechanism}; or (ii) they
provide a small amout of polluting mass. This second possibility is also
unpalatable because a thin layer accreted on the surface of a main sequence
star would be diluted when the stars climb toward the SGB phase by the
extension of the convective envelope. The evidence of large spread in C, N, O,
Na abundances at all luminosities (see Briley, Cohen and Stetson 2002)
strongly argues against a simple contamination of stellar surface.
\item finally the stars more massive than the above range are just the IM-AGB
stars that, at the present, are in our opinion the best candidate polluters.
\end{itemize}
Finally, we must recall that at least half or even 2/3 of stars observed in
globular clusters seem to be heavily affected by large alterations in their
chemical composition. In order to explain this with mass exchange from a RGB
star onto a companion in a binary system, we would probably end up with a huge
number of low-mass X-ray binaries in GCs, at odds with observations.

In summary, our findings and discussion strongly suggest that the polluters
were intermediate mass AGB stars, and $not$ upper RGB stars.

On the other hand, the scenario in which IM-AGB stars are the primary
contributors in shaping the chemical mix of the early cluster environment
still has to face problems (Denissenkov and Herwig 2003, Denissenkov and Weiss
2004, Herwig 2004, Fenner et al. 2004): current models do not seem to
reproduce the required observational pattern. It must be however reminded
that, as noticed by Denissenkov and Weiss, the yields computed from models of
these stars strongly depend on two poorly known physical inputs, namely the
treatment of mass loss and the efficiency of convective transport (see also
Ventura et al. 2002). Further progress in stellar modeling is strongly urged.

\begin{acknowledgements}
{ This research has made use of the SIMBAD data base, operated at CDS,
Strasbourg, France. We thank the ESO staff at Paranal (Chile) for their help
during observing runs, Elena Sabbi for useful discussion on LMXB's in GCs, and
the referee for very careful reading of the manuscript.}
\end{acknowledgements}

\end{document}